\documentclass[%
reprint,
superscriptaddress,
 amsmath,amssymb,
 aps,
]{revtex4-1}

\usepackage{float}
\usepackage{graphicx}
\usepackage{dcolumn}
\usepackage{bm}
\usepackage{inputenc}
\usepackage{gensymb}
\usepackage{lmodern}
\usepackage{makecell}
\usepackage{color}
\usepackage{multirow}

\usepackage{hyperref}
\hypersetup{
colorlinks=true,
linkcolor=blue,
citecolor=blue,
urlcolor=blue
}

\graphicspath{{./images/}}

\usepackage{booktabs}

\usepackage{xcolor}
\usepackage{mathrsfs}

\newcommand{\nm}{\nonumber}

\newcolumntype{M}[1]{>{\centering\arraybackslash}m{#1}}

\begin{document}

\title{
Velocity dependence of kinetic friction by multi-scale \\ Quantum Mechanics/Green's Function molecular dynamics }

\author{Alberto Pacini}
    \email{alberto.pacini3@unibo.it}
    \affiliation{Department of Physics and Astronomy, University of Bologna, 40127 Bologna, Italy}
\author{Seiji Kajita}
    \email{fine-controller@mosk.tytlabs.co.jp}
    \affiliation{Toyota Central R\&D Labs., Inc., 41-1, Yokomichi, Nagakute, Aichi, 480-1192, Japan}
\author{Gabriele Losi}
    \affiliation{Department of Physics, Mathematics and Informatics, University of Modena and Reggio Emilia, 41125 Modena, Italy}
\author{Maria Clelia Righi}
    \email{clelia.righi@unibo.it}
    \affiliation{Department of Physics and Astronomy, University of Bologna, 40127 Bologna, Italy}

\begin{abstract}
Atomistic simulations are powerful tools for investigating tribological phenomena at a fundamental level; however, simulating a tribological system remains challenging due to the multiscale nature of frictional processes. Recently, we introduced a hybrid method, QM-GF, that enables an accurate description of both interfacial chemistry and phononic dissipation in semi-infinite bulks. In this work, we apply this simulation scheme to study the dependence of kinetic friction on sliding velocity. Using a prototypical diamond interface with varying hydrogen coverages, we find that the friction force decreases with increasing sliding velocity, revealing two distinct sliding regimes at low and high speeds. We provide a physical interpretation of this velocity dependence based on the modulation of the frictional force by the sliding motion over the periodic potential energy surface of the interface. High velocities lead to force cancellation, while low velocities result in a net frictional force characterized by a distinctive sawtooth profile.
\end{abstract}

\maketitle

\section{Introduction}
Understanding friction at the fundamental level could play an important role in addressing environmental challenges~\cite{Holmberg-2017}. Friction converts the ordered kinetic energy of two sliding surfaces into disordered thermal energy. This energy degradation occurs at the contacting micro-asperities and is governed by the quantum mechanical interactions between atoms~\cite{bowden2001friction, persson2000sliding, mate2019tribology}. The advent of atomic force microscopy (AFM) and friction force microscopy (FFM) paved the way for the experimental investigation of the elementary processes involved at such scales~\cite{persson2000sliding, mate2019tribology, Ishikawa-2009}. Nonetheless, frictional phenomena remain difficult to probe experimentally due to the challenges in controlling the conditions at the buried, out-of-equilibrium interface. On the other hand, describing non-equilibrium steady states remains a major challenge within theoretical physics~\cite{sasa2005steady}. These difficulties have led to the introduction of phenomenological damping terms to describe energy dissipation, which has been the standard approach adopted in theoretical nanotribology models~\cite{Prandtl-model, Tomlinson-model}. All the complexity arising from the interactions among particles is captured by these phenomenological terms, fitted to experimental data that obey the general theorems of statistical mechanics~\cite{Vanossi-frictionmodelling, Manini2015, Manini_2016, FK-dissipation, braun2013frenkel, gardiner1990handbook}.

Recently, the rapid increase in computational capabilities has made it possible to describe dissipation explicitly by performing molecular dynamics simulations~\cite{Friction-incommensurate, Friction-conslaw, Consoli_2002, Thermolubricity-Franchin, van-den-Ende_2012}. Several studies have reported that energy dissipation depends on the size of the simulated systems relative to the wavelength of the excited phonon modes~\cite{melis2014calculating, kajita2009deep, kajita2010approach, kajita2012simulation}. This dependence has hindered the use of quantum mechanical (QM) calculations, as their computational cost is incompatible with the large system sizes required for such molecular dynamics simulations. QM calculations, however, are crucial for an accurate description of the interfacial potential energy surface (PES) and of the chemical processes occurring under enhanced reactivity conditions, typically found at contacting surfaces~\cite{Kajita-2016, Moseler-2017, Restuccia-2016, Levita-2015, Stella-2017, Peeters-2020, NVT-2018, Kubo-2018, Giulio-Erdemir, Zilibotti-2013, nam-van-tran-2021}.

To overcome these limitations, we recently proposed a new approach that combines QM and Green’s Function (GF) methods. By coupling the interfacial atoms, explicitly treated with quantum mechanics, to a semi-infinite bath of harmonic oscillators represented by the substrates’ GF, it is possible to perform non-equilibrium, quantum-accurate molecular dynamics simulations with a proper energy dissipation mechanism~\cite{GFQM}. In this work, we use this novel QM-GF method to simulate a sliding interface composed of diamond surfaces passivated by hydrogen at different coverages. Our results reveal a friction weakening with increasing sliding velocity. Although the non-monotonic velocity dependence of kinetic friction can be reproduced using phenomenological theories by suitably adjusting the damping parameters, this is, to the best of our knowledge, the first time such behavior is reproduced using first-principles simulations~\cite{Tosatti-friction, F-velocitydependence}. This weakening, which depends on the degree of surface passivation, is characterized by two distinct sliding regimes: \textit{stick-slip} and \textit{continuous} sliding. The two regimes exhibit a marked difference: the stick-slip regime presents a sawtooth profile in the force signal, while the continuous regime shows high-frequency oscillations leading to a net cancellation of the frictional force. The characteristic sliding regime is governed by the interplay between the applied lateral shear and the interfacial shear strength, a static property determined by the electronic behavior of the interface. When the two shears have comparable magnitudes, their competition gives rise to stick-slip, whereas a larger applied lateral shear dominates the interfacial dynamics, leading to the continuous regime.

\section{METHODS}\label{sec:results}

\subsection*{QMGF molecular dynamics in a nutshell}\label{sec:nutshell}

\begin{figure}[htbp]
\centering
\includegraphics[width=\linewidth]{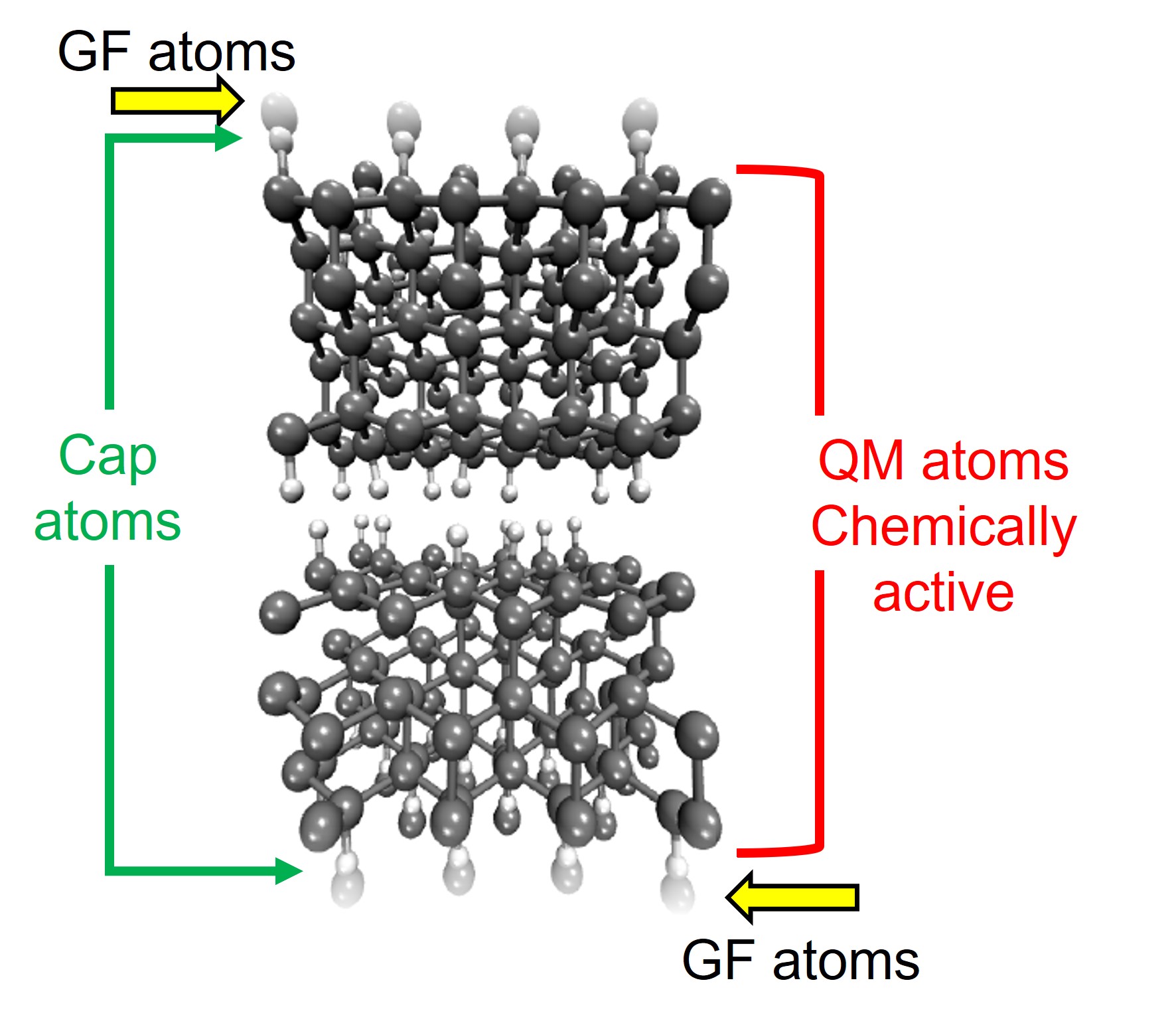} 
\caption{Representation of a frictional interface described by the QMGF scheme. The chemical active region is given by the QM atoms in red, GF atoms representing the semi-infinite bulk where the lateral shear is applied are indicated by yellow arrows, while cap hydrogen atoms are indicated by green arrows.}
\label{pic:Main_system}
\end{figure}

We give a brief explanation of the hybrid QMGF scheme. A full description of the method with details of the algorithms and numerical treatments are presented in~\cite{GFQM}.
Figure~\ref{pic:Main_system} offers a pictorial representation of the hybrid QMGF setup applied to one of the system used in the simulation. The chemically active part of the system consists of the two mated carbon surfaces together with the hydrogen atoms passivating them. The chemical active region is simulated using DFT, explicitly treating the electronic degrees of freedom is necessary to capture quantum effects, such as the Pauli repulsion and the load-enhanced chemical reactivity, which deeply affect the tribological behavior. The boundaries of the interface are connected to \textit{GF atoms} whose dynamics is ruled by the mechanical response of the semi-infinite bulks in terms of the Green's function of an infinite number of harmonic oscillators of first-principles derived spring constants. 
The basic idea of GF MD is, in fact, that all the internal modes of an elastic solid can be integrated out and substituted by effective interactions~\cite{kubo-1966, campana2006practical}. 
The small atoms indicated by green arrows are cap atoms which couple the QM and GF systems. Their positions is constrained by those of QM and GF atoms and the interaction forces are calculated using an add-remove scheme~\cite{swart2003addremove}.
In this way, only the trajectories of the \textit{quantum} and \textit{GF} atoms are needed, and no other bulk atom has to be included in the simulation.

Following the theory of GFMD, three-dimensional displacements of the semi-infinite surface atoms (GF atoms) are described by a linear combination as
\begin{eqnarray}
{\bf u}_p + {\bf u}_g, 
\label{eq:upug}
\end{eqnarray}
where ${\bf u}_p, {\bf u}_g \in {\bf R}^{N\times3}$ are a particular solution and general solution~\cite{StanleyJFarlow} of the equation of motion, while $N$ is the number of the surface atoms in the unit cell. 
The solution ${\bf u}_p$ represents trajectories driven by an external force ${\bf f}$ applied on the surface layer, at initial conditions ${\bf u}_p(t=0) = 0, \frac{d}{dt} {\bf u}_p(t=0) = 0$.
Using the Green's function matrix $A$, the equation of motion can be written as a convolution integral:
\begin{eqnarray}
M \frac{d^2}{dt^2} {\bf u}_p(t) &=& {\bf f}_{\textrm{GF}}(t) \label{eq:psol_main} \nm \\
{\bf f}_{\textrm{GF}}(t) &=& \int_0^{t} A(t-\tau) {\bf f}(\tau) d\tau , \label{eq:covolution}
\end{eqnarray}
On the other hand, the general solution ${\bf u}_g$ describes the evolution of \textit{GF-atoms} without external force but in arbitrary initial conditions ${\bf u}_g(t=0), \frac{d}{dt} {\bf u}_g(t=0)$.
Notably, ${\bf u}_g$ can represent the thermostat and barostat of the system since their statistical fluctuations are related to the Green's function.
In other words, when connecting the GF MD system to the QM system, as in Fig.~\ref{pic:Main_system}(a), ${\bf u}_g$ can be used as a set of parameters to control the temperature and stresses applied to the system.

\subsection*{Details of the Simulation}

We model the interface by mating two C(111) surfaces (most accessible cleavage plane of diamond), with (4×2) in-plane periodicity. The two-faced slabs, each constituted of three bi-layers of carbon atoms, are externally passivated by hydrogen atoms. These atoms represent the cap atoms within the QMGF method and they are externally linked to the GF atoms, representing the bulk response (see fig.~\ref{pic:Main_system}). At the interfacial region, the mating surfaces are passivated with randomly distributed hydrogen atoms using three different coverages: 100\%, 75\% and 50\%. The QMGF molecular dynamics simulations lasted 50 ps at a temperature of 300 K and with 5 GPa of applied vertical load. For each coverage we simulated two different sliding regimes by applying lateral shear stresses of 1 and 5~GPa along the x direction respectively. Within the framework of the hybrid QMGF method, this is equivalent to a situation where we start the friction test by imposing a relative velocity on the semi-infinite solids, as described in \cite{GFQM}.

\begin{figure*}[htbp]
    \centering
    \includegraphics[width=0.9\textwidth]{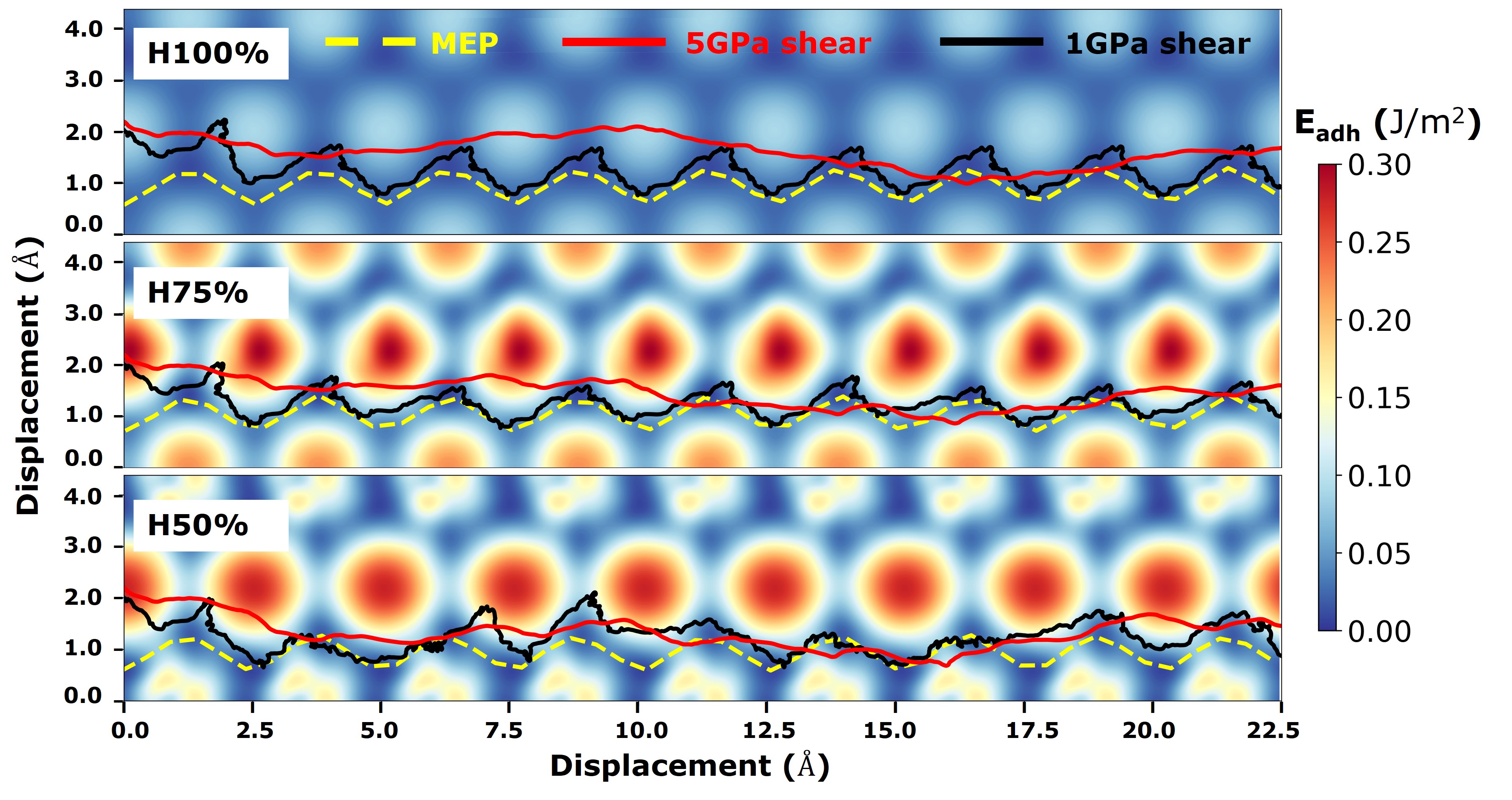}
    \caption{Color maps of the static interfacial PES calculated under 5~GPa of vertical load for 100\% (top), 75\% (middle), and 50\% (bottom) hydrogen passivated systems. Dynamic trajectories at high (5~GPa applied shear, red) and low (1~GPa applied shear, black) sliding velocities are superimposed on the PES together with the MEP (yellow curves). A common energy scale for different PES is used.}
    \label{fig:dynamic-path}
\end{figure*}

\section{RESULTS}\label{sec:results}

\begin{table}[htbp]
\centering
\setlength{\tabcolsep}{3pt}
\renewcommand{\arraystretch}{1.2}

\begin{tabular}{c|c|c|c|c}
\hline
\multicolumn{5}{c}{\textbf{Quantities derived from the sliding dynamics}}\\
\hline
$\theta$ &
$F_{x}^{ext}/A$ &
$\langle v_{x}\rangle$ &
$\langle d_{C}\rangle$ &
$\langle F_{x}^{k}/A\rangle$ \\
(\%) &
(GPa) & (m/s) & (\AA) & (GPa) \\
\hline
\multirow{2}{*}{100} & 5 & 290 & 4.05 & 0.09 \\
 & 1 & 51 & 4.04 & 0.15 \\
\hline
\multirow{2}{*}{75} & 5 & 285 & 3.76 & 0.18 \\
 & 1 & 51 & 3.72 & 0.21 \\
\hline
\multirow{2}{*}{50} & 5 & 281 & 3.28 & 0.27 \\
 & 1 & 48 & 3.25 & 0.27 \\
\hline
\end{tabular}

\caption{Results of the QM-GF MD simulations. For each hydrogen coverage $\theta$ and applied lateral force per unit area $F_{x}^{ext}/A$ (or equivalent sliding velocity $\langle v_{x}\rangle$), we report the surface separation $\langle d_{C}\rangle$ and the kinetic friction force per unit area $\langle F_{x}^{k}/A\rangle$. All values are time averages calculated over the QMGF simulations.}
\label{tab:mu_table}
\end{table}

Tab.~\ref{tab:mu_table} reports the kinetic friction forces and interfacial separations at low and high sliding velocities for each degree of hydrogen coverage. These quantities are calculated as averages over the MD simulation time. In particular, the kinetic friction force, $\langle F_{x}^{k}\rangle$, is obtained as the time average of the $x$-component of the total force applied to the \text{GF-atoms}. These are precisely the forces appearing in the convolution integral of Eq.~\ref{eq:covolution}. The values in Tab.~\ref{tab:mu_table} reveal a weakening of kinetic friction with increasing sliding velocity. This velocity dependence gradually diminishes as the hydrogen coverage decreases and disappears completely for the half-covered interface.

To further analyze the results of the dynamic QM-GF simulations, we superimposed the trajectories followed by the systems during sliding on top of the potential energy surface (PES) of the interface (Fig.~\ref{fig:dynamic-path}). The PES is a static property of the interface describing the adhesion energy between the two surfaces as a function of their relative lateral position, $\gamma\left( x,y \right)$. The color maps in Fig.~\ref{fig:dynamic-path} display the PESs of the passivated interfaces under a vertical load of 5~GPa. Dashed yellow lines indicate the minimum energy path (MEP) of the PES, representing the path followed by the system under adiabatic conditions. We identify two distinct sliding regimes. At high sliding velocity (5~GPa applied shear, red curves), the trajectories deviate from the MEPs and cross higher energy regions of the PES, whereas at lower sliding velocity (1~GPa applied shear, black curves), the trajectories closely follow the MEPs.

The gradient of the PES along the trajectories represents the shear stress of the interface during sliding~\cite{Zilibotti-2011}:
\begin{equation} \label{eq:tauPES}
\tau_{{\textrm PES}} \equiv -\nabla\gamma\left( x, y\right)\big|_{{\textrm trajectory}} \nm
\end{equation}

It is important to note that the trajectories in Fig.~\ref{fig:dynamic-path} and the $\tau_{PES}$ defined in Eq.~\ref{eq:tauPES} correspond to the center-of-mass degree of freedom of the surfaces, which results from the collective dynamics of the atomic degrees of freedom. Indeed, in previous work, we showed that the onset of frictional slip is a nucleation process, where atomic displacements along the PES are not coherent~\cite{Reguzzoni-2010}.

\begin{figure}[htbp]
    \centering
    \includegraphics[width=\linewidth]{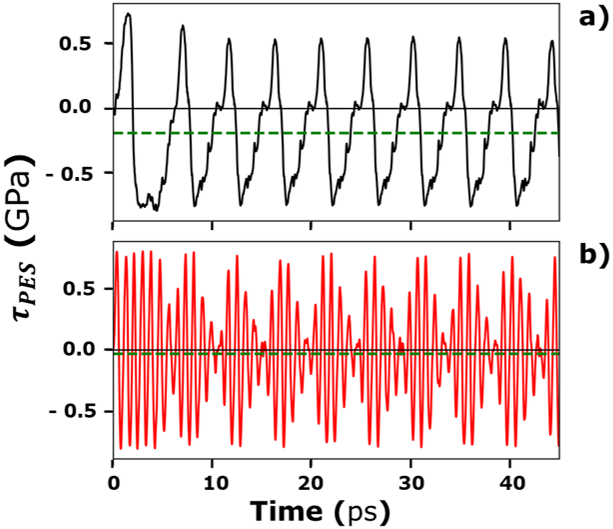}
    \caption{PES-induced forces per area along the direction of sliding for the 100\% hydrogenated interface. The black curve represents \textit{stick-slip} regime (1~GPa applied shear) while the red one is \textit{continuous sliding} (5~GPa applied shear). The green dashed lines represent the time averages of the signals.}
    \label{fig:Stick-slipVScontinuous}
\end{figure}

Figure~\ref{fig:Stick-slipVScontinuous} shows the $x$-component of $\tau_{{\textrm PES}}$ for the fully hydrogenated interface. The regularity of the force oscillations in Fig.~\ref{fig:Stick-slipVScontinuous} indicates the presence of a characteristic frequency in the force signal, known as the washboard frequency. This frequency arises from the sliding motion at constant velocity over a periodic potential and can be estimated as:
\begin{equation}
\nu_{{\textrm wash}} = \frac{v_{\textrm {slide}}}{a} \nm
\end{equation}
where $v_{\textrm {slide}}$ is the sliding velocity and $a$ is the lattice parameter. We find $\nu_{{\textrm wash}}^{1,{\textrm GPa}} = 0.19~\text{THz}$ and $\nu_{{\textrm wash}}^{5,{\textrm GPa}} = 1.12~\text{THz}$ for the two sliding regimes (Fig. 2 in SI).

The shear stress signals in Fig.~\ref{fig:Stick-slipVScontinuous} also exhibit clear differences in their shape profiles. The sawtooth pattern, characteristic of the stick-slip regime, appears in the signal at lower sliding velocity (panel a). In this regime, there is a competition between the applied lateral force and the interfacial forces determined by the corrugation of the PES. The system moves from one PES minimum to the next by climbing over the saddle points. During the ascent of the PES barrier, the system is slowly displaced from the minimum and accumulates elastic energy. This corresponds to the stick phase, where the PES-derived force is negative and lasts longer (negative half-waves in Fig.~\ref{fig:Stick-slipVScontinuous}). Once the saddle point is reached, the system enters the slip phase. The PES-induced force becomes positive, pushing the system toward the next minimum, but the duration of this phase is shorter (positive half-waves in Fig.~\ref{fig:Stick-slipVScontinuous}). These non-adiabatic jumps create an asymmetry between the durations of the positive and negative half-waves, resulting in a net negative average friction force, indicated by the green dashed line.

When the sliding velocity increases, the external shear becomes the dominant driver of interfacial dynamics (Fig.~\ref{fig:Stick-slipVScontinuous}b), and a continuous sliding regime emerges. In this regime, the interface is driven over the PES by the external shear at high velocity, leaving no time to relax along the MEP. The stick time becomes comparable to the slip time, and the force oscillations become more symmetric. This leads to a decrease in the magnitude of the negative average force, corresponding to the reduction in kinetic friction observed in Tab.~\ref{tab:mu_table}.
\section{Conclusion} \label{sec:conclusion}
In this work, we employed the novel GF/QM method to gain insights into the fundamental nature of kinetic friction. The realistic treatment of energy dissipation mechanisms through the bulk stabilizes the interfacial dynamics in out-of-equilibrium simulations, enabling quantitative comparisons of kinetic friction coefficients at the nanoscale. We observe a weakening trend of kinetic friction with increasing sliding velocity. By combining the results of dynamic and static simulations, we identify two distinct sliding regimes governed by the magnitude of the applied shear. At low applied shear, the system trajectory during sliding closely follows the MEP—the equilibrium path—resulting in a \textit{stick-slip} motion. At higher applied shear, the system enters a \textit{continuous} sliding regime, in which the trajectory deviates significantly from the MEP and follows an almost linear path across the PES. In the \textit{stick-slip} regime, the competition between the applied lateral shear and the interfacial shear derived from the PES produces a characteristic sawtooth profile associated with a larger net frictional force. As the applied lateral shear increases, the influence of the PES on interfacial dynamics becomes less significant, and continuous sliding emerges, characterized by substantial force cancellation and a consequent reduction in friction signal. These results highlight the great potential of the QM-GF method to provide highly accurate insights into interfacial phenomena under non-equilibrium conditions. This approach may pave the way for the investigation of other multiscale processes in which the infinite number of bulk degrees of freedom—typically neglected in \textit{ab initio} molecular dynamics—plays a crucial role in determining the system’s response to external stimuli.

\begin{acknowledgments}
These results are part of the ”Advancing Solid Interface and Lubricants by First Principles Material Design (SLIDE)” project that has received funding from the European Research Council (ERC) under the European Union’s Horizon 2020 research and innovation program (Grant agreement No. 865633). We thank Dr. C. Cavazzoni for the help in implementing the QMGF MD method within the cp.x code. The pictures in the present paper are created with the help of VMD~\cite{VMD}, and Matplotlib~\cite{Hunter-2007}. 
\end{acknowledgments}

\section*{Ethics declarations}
The Authors declare no Competing Financial or Non-Financial Interests

\section*{Data Availability}
All data were available from the corresponding authors upon reasonable request.

\section*{Code Availability}
The related codes are available from the corresponding authors upon reasonable request.

\section*{Author Contributions}
The research was conceived by MCR and SK, supervision and project administration by MCR. The Green's Function subroutine was developed by SK and Nobuaki Kikkawa. Its linking with AIMD was implemented by MCR's group and SK. The simulations were carried out by AP and GL. All authors discussed the results and contributed to writing the manuscript.



\bibliographystyle{apsrev4-1}
\bibliography{biblio}


\end{document}